\documentclass[pra,preprint,onecolumn]{revtex4}%
\usepackage{amsmath}
\usepackage{graphicx}
\usepackage[latin9]{inputenc}
\usepackage[T1]{fontenc}
\usepackage[english]{babel}
\usepackage{amsfonts}
\usepackage{amssymb}

\newcommand{\be}{\begin{equation}}
\newcommand{\ee}{\end{equation}}
\newcommand{\ben}{\begin{eqnarray}}
\newcommand{\een}{\end{eqnarray}}

\begin{document}
\title{Memory endowed US cities and their demographic interactions}
\author{A. Hernando}
\email{alberto.hernandodecastro@epfl.ch}
\affiliation{Laboratory of Theoretical Physical Chemistry, Institut des Sciences et
Ing\'enierie Chimiques, \'Ecole Polytechnique F\'ed\'erale de Lausanne,
CH-1015 Lausanne, Switzerland}
\author{R. Hernando}
\affiliation{Social Thermodynamics Applied Research (SThAR), Madrid, Spain}
\author{A. Plastino}\email{plastino@fisica.unlp.edu.ar}
\affiliation{National University La Plata, Physics Institute (IFLP-CCT-CONICET)  La Plata, Argentina}
\affiliation{Universitat de les Illes Balears and IFISC-CSIC, Palma de Mallorca, Spain}
\author{E. Zambrano}
\affiliation{Laboratory of Theoretical Physical Chemistry, Institut des Sciences et
Ing\'enierie Chimiques, \'Ecole Polytechnique F\'ed\'erale de Lausanne,
CH-1015 Lausanne, Switzerland }

\date{\today}

\begin{abstract}
A quantitative understanding of cities' demographic dynamics
is becoming a potentially useful tool for planning sustainable
growth. The concomitant theory  should reveal details of the
cities' past and also of its interaction with nearby urban
conglomerates for providing a reasonably complete
picture. Using the exhaustive database
of the Census Bureau in a time window of 170 years, we exhibit
here empirical evidence for time and space correlations in the
demographic dynamics of US counties, with a characteristic memory-time of 25 years
and typical distances of interaction of 200 km. These correlations are
much larger than those observed in an European country (Spain),
giving to the US a more coherent evolution.
We also measure the resilience of US cities to historical events,
finding a demographical {\it post-traumatic amnesia} after
wars (as the Civil War) or economic crisis (as the 1929 Stock Market Crash).\\
\end{abstract}
\maketitle

\section{Introduction}

A half of the human population lives in urban areas \cite{pop}.
Asking whether the present population's growth rates are
economically and ecologically  sustainable is a recurrent question
\cite{pop2} that justifies efforts directed towards the
development of quantitative unified theories of urban living
\cite{betwest}.
A countless number of degrees of freedom is involved in  a city's
evolution, host of  individual contributions, involving  millions
of people, acting on their own free will. Devising a unified
theory constitutes a formidable challenge.
However, despite this intrinsic difficulty, many advances
have been made in the last years. During the XX century
many regularities were reported, such as Zipf's Law in the
city-population rank distribution
\cite{zipf,mbat2,power,X1,nosPLA09}, or the celebrated Gibrat's
Law of proportional growth applicable to cities
\cite{gibrat0,gibrat,city1,city2,nosJRSI13,nosJRSI14}. In addition,
empirical data show that the scaling with population of internal
degrees of freedom in an urban body such as i) the structure of
road networks or urban sprawl patterns \cite{gibrat,diff,roads2};
and ii) metrics such as wages or crime rates \cite{lbet2,PnasW},
both follow predictable tendencies that can be mathematically
described. In addition to a city's growth, electoral processes and
many other social phenomena have been successfully  modelled as
well  \cite{santo,firms,1sta,oppi,power,nosPLA09,nosEPJB10}. Also,
collective modes emerge when cities are considered as entities
that evolve and interact in a coherent fashion. The interaction
between cities \cite{nosJRSI14} (as measured by, for instance, by
the number of crossed phone calls \cite{gmodel} or human mobility
\cite{mob1}) displays predictable characteristics. Indeed, an
analogy between the evolution of the population of an ensemble of
cities and the random movement of particles in a fluid was made by
conjoining Gibrat's Law of proportional growth with  Brownian
motion. In such an approach the size $X$ at time $t$ of the $i$-th
geometrical Brownian walker follows the dynamical
equation\cite{gibrat0,city1,city2,gibrat,nosEPJB12b,nosJRSI13}
\begin{equation}
\dot{X}_i(t) = X_i(t) v_i(t),
\end{equation}
where the dot denotes a time-derivative and $v_i(t)$ is the growth
rate, which is described as a Wiener coefficient with covariance
$\mathrm{Cov}[v_i(t),v_j(t')]=\sigma_v^2\delta_{ij}\delta(t-t')$
---$\delta$~standing for the delta function and $\sigma_v$ for the
standard deviation of the growth rates--- i.e., uncorrelated and
memoryless dynamics. Considering a new variable defined as
$u_i(t)=\ln X_i(t)$ \cite{nosEPJB12b}, we recover all the
properties of the physical Ideal Gas in what one may call the
scale-free ideal gas \cite{nosPA10,nosJRSI13}. Exhaustive
empirical observations of the dynamics of Spain's population
demonstrates that this analogy can be used to formulate a
Thermodynamics of population flows \cite{nosPRE12}. Moreover, we
have recently shown that this assumption of uncorrelated evolution
fails for cities that are close neighbors \cite{nosJRSI14}.
Additionally, the evolution of cities-population exhibits memory,
indicating that we deal with a non-Markovian process. These
results are indicative of i) a rich and complex phenomenology
underlying population flows, and ii) that our models should go
beyond the Ideal Gas to include pairwise interactions and inertia,
as in the case of  real gases in Statistical Physics. The presence
of some kind of correlations has been also independently proposed
in Ref. \cite{mbat2}, where it is shown via numerical experiments
that it is necessary to introduce conditioned sampling to
reproduce Zipf's Law, simulating what the authors call
`coherence'. These advances, both in the internal structure and
the functionality of cities, as well as in the properties of an
ensemble of interacting cities, encourage the searching for a
unified theory.

\section{Results}

\subsection{Stochastic properties of US growth rates}

We look for quantitative space-time patterns underlying US
demographics and ascertain which trends are universal and which
are local by comparison with Ref. \cite{nosJRSI14} for Spanish
cities. An exhaustive analysis of the US population is made using
the Census Boreau database of counties' population \cite{census}.
We have used data from 1830 to 2000 (170 years), in a time window
that covers relevant historical events such as the Civil War, both
World Wars, and the 1929 Stock Market Crash. More than 3000
counties (all of them with available data) are considered in our
study, whose spatial distribution is depicted in Fig.~1. We
verified the validity of Gibrat's Law, including the correction
for smaller populations, discovered in Ref. \cite{nosJRSI13}. To
this effect we added a new term to the proportional law,  a
'finite size noise' (FSN) of the form
\begin{equation}
\dot{X}_i(t) = X_i(t) v_i(t) + \sqrt{X_i(t)}w_i(t),
\end{equation}
where $w_i(t)$ is an independent Wiener coefficient. This term is
a direct consequence of the Central Limit Theorem, as shown in
Refs. \cite{nosJRSI13,nosPRE12}, due to the independent nature of
the $w_i(t)$. The variation of the population $X_i$ is much
smaller than the variation of the growth rates. Thus, for the
later, the standard deviation  $\sigma_{\dot{x}}$ in our  time
windows becomes
\begin{equation}
\sigma_{\dot{x}}(X_i) \cong X_i \sigma_v + \sqrt{X_i}\sigma_w,
\end{equation}
where $\sigma_v$ and $\sigma_w$ are the deviations of the
proportional and finite size noise, respectively. Results are
displayed in Fig. 1. It is clear that the median of the
deviations, as a function of the county population, follows 1) a
FSN trend for $X<35000$ inhabitants, and 2)  proportional growth
for $X>35000$. Remarkably enough, a linear fit to the log-log
representation
---where power laws become straight lines with slope-values linked to
 the exponents--- gives for the exponents $0.508\pm0.025$
(with $\log(\sigma_w)=2.6\pm0.2$) for the first trend and
$1.04\pm0.02$ (with $\log(\sigma_v)=-3.1\pm0.2$) for the second
one, with a coefficient of determination $R^2$ of $0.96$ and
$0.994$, respectively. Since these values  almost coincide with
the expected ones ($1/2$ and $1$), we can regard the Gibrat plus
FSN Law as verified, a rather significant observation.

\subsection{Measuring memory}

To check whether US counties exhibit  memory effects, we appeal to
the Pearson's product-moment correlation coefficient
$c_y(t)=\mathrm{Corr}[\dot{X}(y),\dot{X}(t)]$, using as samples
the list of counties' growth rates for i) the year $y$ and ii) any
precedent time $t$ for which data are available.
%So as to discard effects
%related to the total population growth, we use from now on the
%relative populations $x_i(t)=X_i(t)/N(t)$, where $N(t)$ is the
%total US population at any given year.
We  find that the averaged time-correlation over $n_y=12$ samples
(from 1890 to 2000), defined as $\langle c(\Delta
t)\rangle=n_y^{-1}\sum_y c_y(y+\Delta t)$ and calculated for
consecutive census instantiations ($\Delta t=10$ years), shows a
behavior similar to that found for  Spanish cities
\cite{nosJRSI14}: large cities exhibit greater inertia. We can
attribute  the smaller counties'  loss of memory  to the FSN term,
which becomes important for them (see Fig. 2a).
%Considering now populations larger than $10^5$ inhabitants,
For larger intervals of time $\Delta t$, we find a clear decay of
the averaged time-correlation. Remarkably enough, the correlations
are much larger than those found for  Spanish cities (Fig. 2b).
Considering only the first 50 years, a fit to an exponential decay
gives us a characteristic time of $25\pm7$ years ($R^2=0.990$),
but surprisingly enough, the correlation eventually becomes
negative after $\sim 60$ years.

In order to  gain a deeper understanding of this unexpected trend,
and also to check whether it is caused by a non-homogeneous
behavior of the correlations, we have independently studied all
the contributions $c_y(t)$ for several years $y$ (and precedent
times $t$). We find that, although for all $y$ a decay of the
correlation with time is always present, historical events clearly
modulate these correlations. For the growth rates from 1890 to
1950, we find that, irrespective of the year, no memory remains in
the demographics of US counties from the years that precede the
Civil War, in a kind of `post-traumatic amnesia'. For the second
half of the XX century, we find, in general, a slower decay
---larger memory--- than for other time-periods. The most
important historical event that one immediately detects (by simple
inspection) regarding the cities' memory  is the economic crisis
after the 1929 Stock Market Crash. Again, irrespective of the
year, one still encounters (i) a correlation's fall and (ii)  loss
of  memory regarding precedent decades. Thus, instead of an
homogeneous year-independent decay of the correlations with time,
we find a decay with a strong dependence on historical events.
However, some unanswered questions remain
---as 1) why do the years 1940 and 1950 seem to be not much  afflicted
by  amnesia? or 2) the reason for the  strong fluctuations between
the years 1990 and 2000. Accordingly, more research along this
line of `quantitative History' remains to be undertaken.

\subsection{Measuring interactions}

We consider now  spatial correlations. The pairwise Pearson
product-moment correlation coefficient
$C_{ij}=\mathrm{Corr}[\dot{X}_i,\dot{X}_j]$ of the $i$ and $j$-th
counties is obtained using as samples the evolution of the growth
rates of each county in a given  time window. We speak here of the
XX century, from 1900 to 2000 (10 sample sets). We compare the
value obtained per each pair with the distance between counties
$d_{ij}$. The averaged value ---obtained as $C(d) =
\sum_{ij}C_{ij}\delta(d-d_{ij})/\sum_{ij}\delta(d-d_{ij})$---
exhibits a clear dependence on distance, demonstrating the
entanglement between US populations-nuclei. The tail of the decay
displays a long-range behavior (see Fig. 4), and the pertinent
curve can be nicely fitted to an analytical expression of the form
\begin{equation}
C(d) = \frac{C(0)}{1+|d/d_0|^\alpha},
\end{equation}
with $C(0)=0.62\pm0.02$, $d_0=215\pm32$ km, and
$\alpha=0.71\pm0.04$, for a $R^2$ coefficient of $0.997$.
Remarkably, the correlations are much larger that those observed
for Spain \cite{nosJRSI14}, with a larger characteristic distance
(215 US vs. 80 km for Spanish cities) and a much slower decay
($\alpha=0.71$). Note that for an  inverse-square law, $\alpha=2$.

The comparison between USA and Spain is illustrated by Fig. 4.
Results confirm the conjecture that US cities evolve in a more
coherent fashion that the cities in Spain, notwithstanding the
fact that the US surface is 20 times larger than Spain's, while
its population is 6.5 times larger. We may speak of an
integration-coherence for US cities that seems to be lacking in
Europe, as has also been proposed in Ref. \cite{mbat2}: Zipf's Law
emerges when the largest US cities are considered but not when
this is done on a state-by-state basis, whereas in Europe, Zipf's
law emerges for each country as a whole, and not when all the
European continent is considered. The standard deviation observed
for the US is of $\sim0.4$ and does not change with distance. The
expected theoretical width for a bivariate normal distribution for
the same number of samples is $1/3$ \cite{biv}, a 20\% smaller
than the measured one. Thus, we gather that additional factors are
involved in the US pairwise correlation. One can attribute to the
distance-factor an 80\% of the city-city entanglement. We expect
that some of these extra contributions could be associated to
local factors, such as the transportation network, the particular
socio-economical status of the city, and/or special historical
links between some population-nuclei. A detailed analysis of the
pairwise correlations of a selected county --instead of the
coarse-grained viewpoint adopted  here--- when crossed with other
relevant metrics, may help to gain a deeper understanding of the
particular demographic and/or economic status, present and future,
of a given urban area.

\section{Summary and conclusions}

Demographic US patterns display a rich and complex phenomenology,
including both space and time correlations. US cities exhibit a
strong link with their past. In an exercise of quantitative
History, we have found that relevant historical events, such as
the Civil War and the 1929 economic crisis, leave a strong imprint
in the demographic dynamics, that one may call a `post-traumatic
amnesia'. The spatial correlations, much larger than those
observed in Europe, indicate a high level of coherence and suggest
that the evolution of any single city can not be understood
without taking the whole collective of cities into account. We
feel that these empirical findings are relevant to understanding
the country at a collective macroscopic level. Also, some
microscopic insights are gained that may help city planners to
improve their panoply of tools \cite{census,tools2}.

%%%%%%%%%%%%%%Fig 1%%%%%%%%%%%%%%%%%%%%%%%%%%%
\begin{figure}[t!]
\begin{center}
{\bf A)}\hfill\hbox{}\\
\centerline{\includegraphics[width=0.7\linewidth,clip]{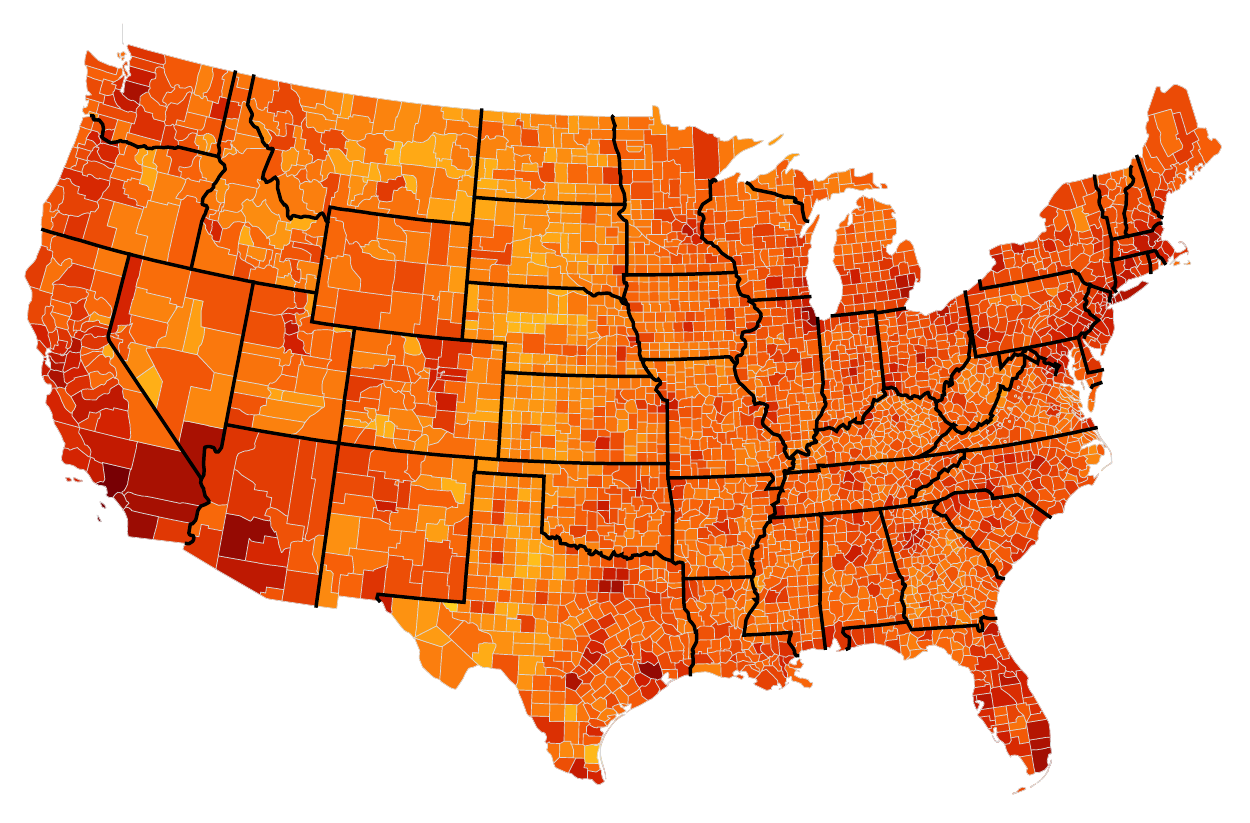}}
{\bf B)}\hfill\hbox{}\\
\centerline{\includegraphics[width=0.7\linewidth,clip]{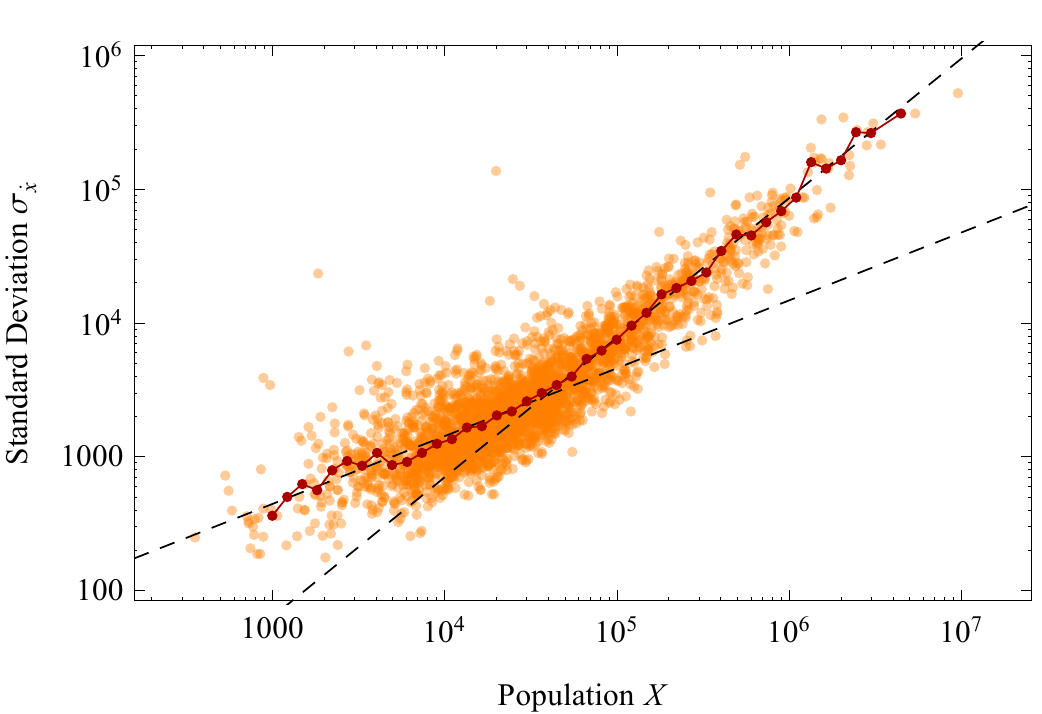}}
\caption{{\bf A)} spatial distribution \cite{map} and population of US counties \cite{census} (the darker, the larger population).
{\bf B)} Standard deviation $\sigma$ of county's growth rate as a
function of the total population $X$ (orange dots). The median
(red line and dots) clearly follows the proportional Gibrat's law
for populations larger than $35000$ inhabitants (slope 1 in a
log-log representation), while it follows the expected trend of
finite size noise for smaller populations (slope 1/2). Dashed
lines follow the linear fit performed per each contribution (see
text). }\label{fig1}
\end{center}
\end{figure}
%%%%%%%%%%%%%%%%%%%%%%%%%%%%%%%%%%%%%%%%%
%%%%%%%%%%%%%%Fig 2%%%%%%%%%%%%%%%%%%%%%%%%%%%
\begin{figure}[t!]
{\bf A)}\hfill\hbox{}\\
\centerline{\includegraphics[width=0.7\linewidth]{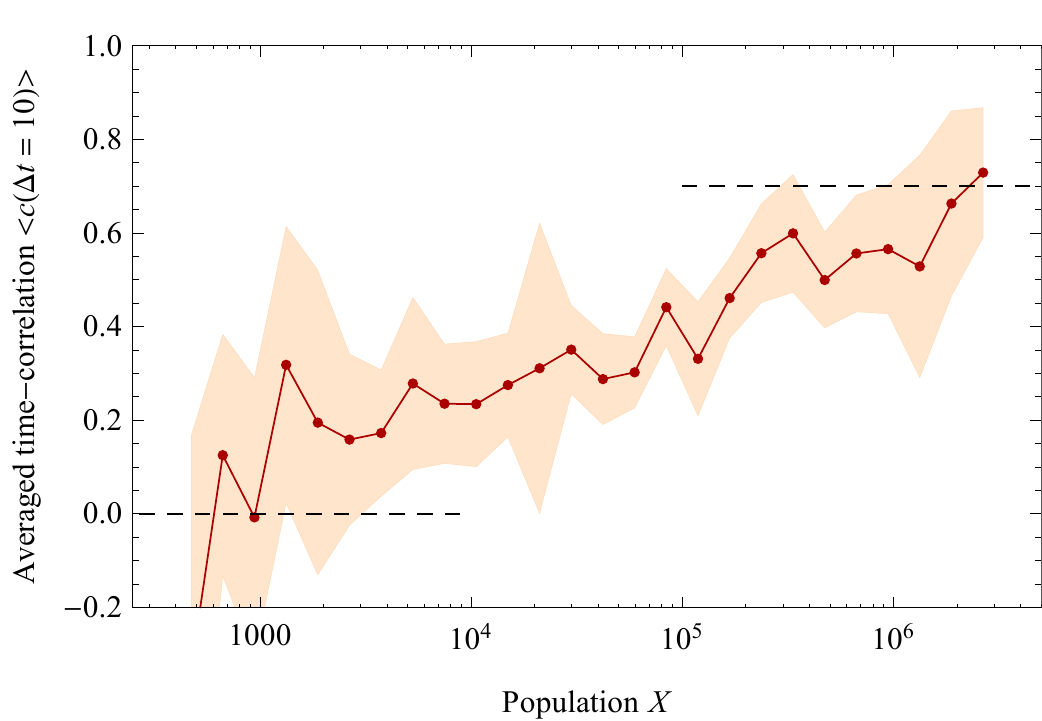}}
{\bf B)}\hfill\hbox{}\\
\centerline{\includegraphics[width=0.7\linewidth]{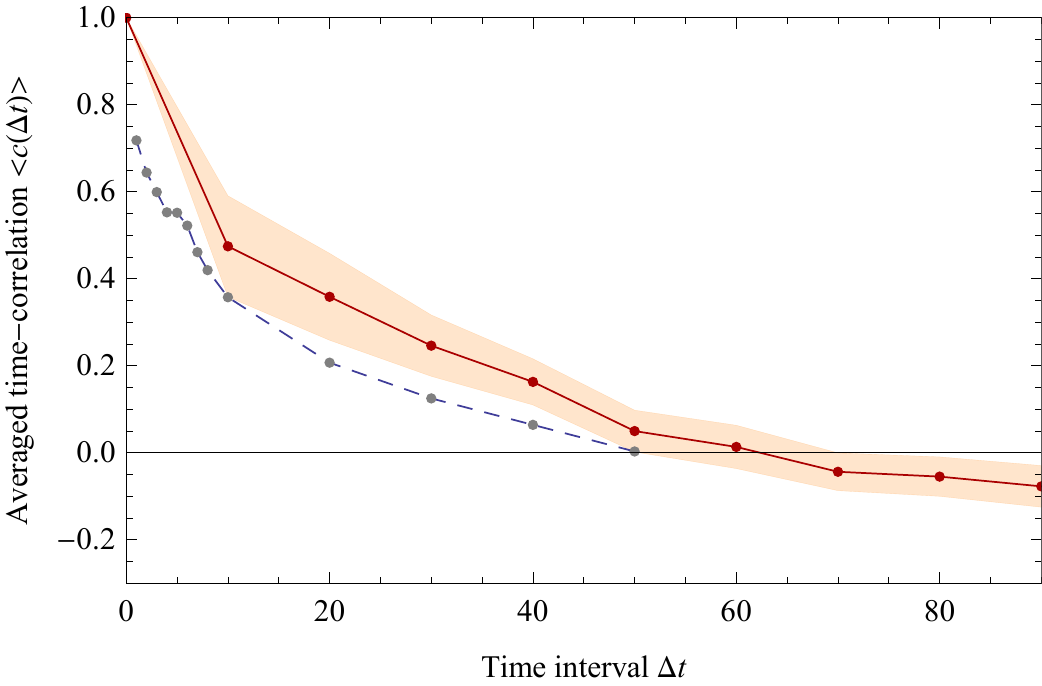}}
\caption{Averaged time-correlations in population growth. {\bf A)}
Averaged time-correlations after 10 years $\langle c(\Delta
t=10)\rangle$ versus town-sizes $X$ (dot-line). The shaded area
represents the width determined by one standard deviation.
Horizontal dashed lines are asymptotic levels
to follow the decrease of the correlation at lower populations.
{\bf B)} Averaged time-correlation $\langle c(\Delta t)\rangle$
for the relative growth of towns populated by more that 10,000
inhabitants. Shaded areas represent  widths determined by one
standard deviation. The time-correlation for Spanish cities is
also shown for comparison's sake (grey line and dots).
}\label{fig2}
\end{figure}
%%%%%%%%%%%%%%%%%%%%%%%%%%%%%%%%%%%%%%%%%
%%%%%%%%%%%%%%Fig 3%%%%%%%%%%%%%%%%%%%%%%%%%%%
\begin{figure}[t!]
{\bf A)}\hfill\hbox{}\\
\centerline{\includegraphics[width=0.7\linewidth]{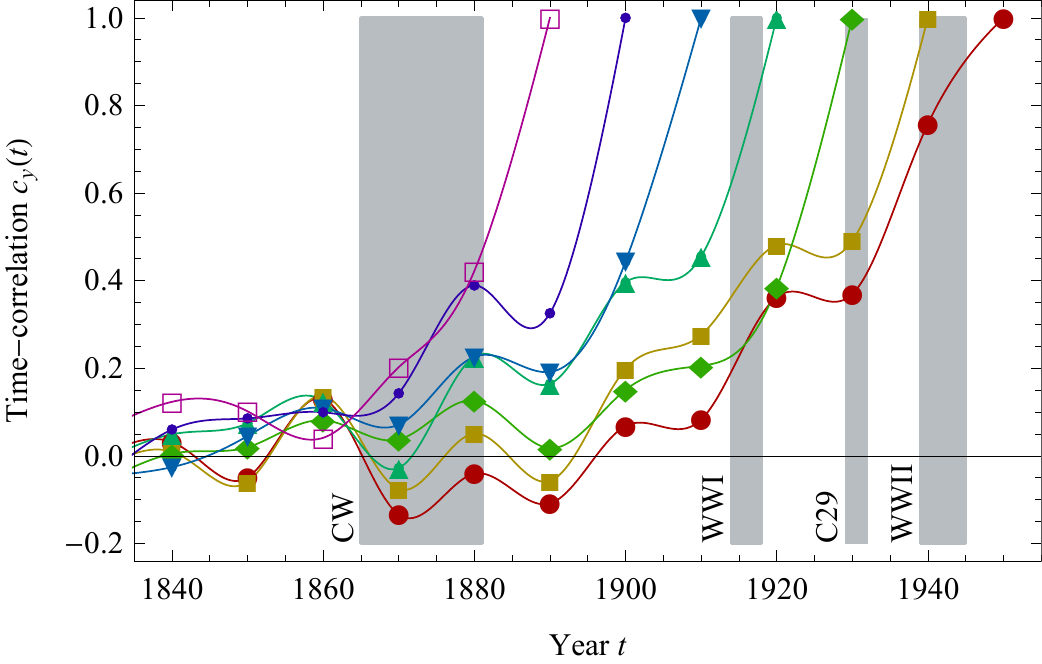}}
{\bf B)}\hfill\hbox{}\\
\centerline{\includegraphics[width=0.7\linewidth]{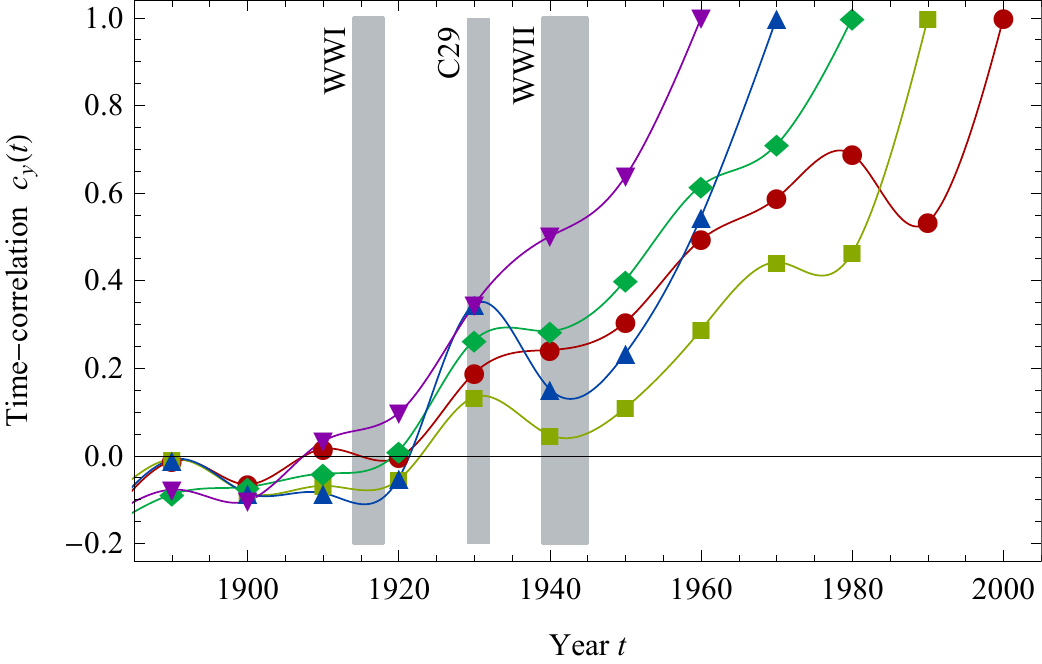}}
\caption{Time-correlations in population growth $c_y(t)$ of the
year $y$ with all previous years denoted by $t$. Historical events
as the Civil War (CW), the First World War (WWI), the 1929 Stock
Market Crash (C29), and the Second World War (WWII), are marked
with shadows. {\bf A)} Correlations for years $y=1950$ (red
circles), 1940 (yellow squares), 1930 (green diamonds), 1920 (turquoise
up-triangles), 1910 (blue down-triangles), 1900 (navy small
circles), and 1890 (purple empty squares). In all cases, the
correlation becomes almost zero after the Civil War. {\bf B)}
Correlations for the last half of XX century: years $y=2000$ (red
circles), 1990 (yellow squares), 1980 (green diamonds), 1970 (blue
up-triangles), and 1960 (purple down-triangles). A general pattern
of decay is apparent for all years, and remarkably, the five
curves drop after the Stock Market Crash of 1929. }\label{fig2}
\end{figure}
%%%%%%%%%%%%%%%%%%%%%%%%%%%%%%%%%%%%%%%%%
%%%%%%%%%%%%%%Fig 4%%%%%%%%%%%%%%%%%%%%%%%%%%%
\begin{figure}[t!]
\begin{center}
\centerline{\includegraphics[width=0.7\linewidth,clip]{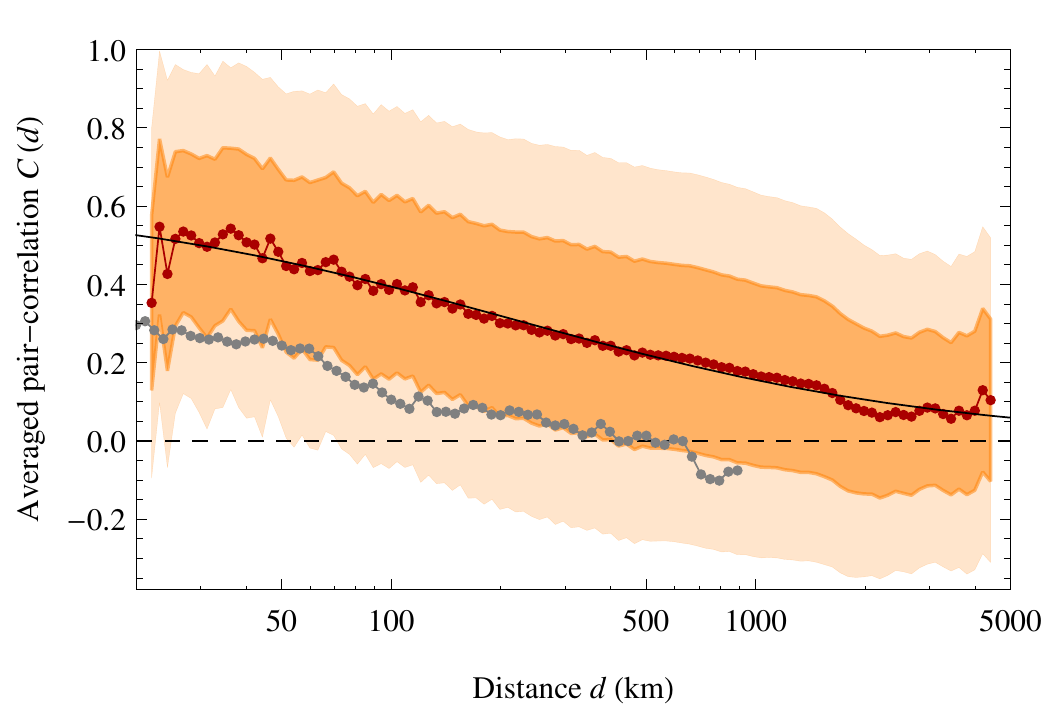}}
\caption{Pairwise correlation $C$ versus distance between counties
$d$. The mean value for a given distance is indicated (red line
and dots) together with one standard deviation (darker shadow) and
by two of them (lighter shadow). The black line follows the
analytical fit of Eq. (4). The spatial correlation for Spanish
cities is also shown for comparison's sake (grey line and dots).
}\label{fig4}
\end{center}
\end{figure}
%%%%%%%%%%%%%%%%%%%%%%%%%%%%%%%%%%%%%%%%%


\begin{thebibliography}{99}

\bibitem{pop} UN-Habitat, {\it State of the World's Cities 1010/2011 --- Cities for All: Bridging the Urban Divide} (2010) www.unhabitat.org

%1
\bibitem{pop2} Schellnhuber HJ, Molina M, Stern N, Huber V, Kadner S. {\it Global Sustainability: A Nobel Cause} (Cambridge Univ. Press, 2010)

%2
\bibitem{betwest} Bettencourt L, West G. A unified theory of urban living {\it Nature} {\bf 467}, 912  (2010)

%3
\bibitem{mbat2} Cristelli M, Batty M, Pietronero L. There is more than a Power Law in Zipf {\it Scientific Reports} {\bf 2}, 812 (2012)

%\bibitem{bat1} Batty M (2008) The Size, Scale, and Shape of Cities, {\it Science} 319:769.

%4
\bibitem{zipf} Zipf GK. \emph{Human Behavior and the Principle of Least Effort} (Addison-Wesley, Cambridge, MA, 1949).

%5
\bibitem{power} Newman MEJ. Power laws, {Pareto} distributions and {Z}ipf's law. {\it Contemp Phys} {\bf 46}, 323 (2005)

%6
\bibitem{X1} Baek SK, Bernhardsson S, Minnhagen P Zipf's law unzipped, {\it New J Phys} {\bf 13}, 043004  (2011) .

%7
\bibitem{nosPLA09} Hernando A, Puigdom\`enech D, Villuendas D, Vesperinas C, Plastino A. Zipf's law from a Fisher variational-principle, {\it Phys Lett A} {\bf 374}, 18 (2009)

%8
\bibitem{gibrat0} Gibrat R  {\it Les In\'egalit\'es \'economiques}, (Librairie du Recueil, Sirey, Paris, 1931).

%9
\bibitem{gibrat} Rozenfeld H, Rybski D, Andrade JS, Batty M, Stanley HE, Makse HA (2008) Laws of Population Growth, {\it Proc Natl Acad Sci USA} 105:18702.

%10
\bibitem{city2} Gabaix X, Ioannides YM. \emph{Handbook of Regional and Urban Economics}, Vol. 4 (North-Holland, Amsterdam, 2004).

%11
\bibitem{city1} Blank A, Solomon S.  Power laws in cities population, financial markets
and internet sites (scaling in systems with a variable number of components), {\it Physica A} {\bf 287}, 279 (2000)

%12
\bibitem{nosJRSI13} Hernando A, Hernando R, Plastino A,Plastino AR. The workings of the maximum entropy principle in collective human behaviour, \emph{J R Soc Interface} {\bf 10}, 20120758  (2013)


%28
\bibitem{nosJRSI14} Hernando A, Hernando R, Plastino A. Space-time correlations in urban sprawl, \emph{J R Soc Interface} {\bf 11}, 20130930  (2013)

%13
\bibitem{diff} Makse HA, Andrade JS, Batty M, Havlin S, Stanley HE. Modelling urban growth patterns with correlated percolation, {\it Phys. Rev. E}, {\bf 58}, 7054 (1998)

%\bibitem{roads1} Strano E, Nicosia V, Latora V, Porta S, Barthelemy M (2012) Elementary processes governing the evolution of road networks, {\it Scientific Reports}, {\bf 2}, 296.

%14
\bibitem{roads2} Masucci AP, Stalinov K, Batty M. Limited Urban Growth: London's Street Network Dynamics since the 18th Century, {\it PLoS One}, {\bf 8}, e69469  (2013)

%\bibitem{roads3} Barthelemy M, Bordin P, Berestycki H, Gribaldi M (2013) Self-organization versus top-down planning in the evolution of a city, arXiv:1307.2203.

%15
\bibitem{lbet2} L. Bettencourt. {\it Science} {\bf  340}, 1438 (2013).

%16
\bibitem{PnasW} Bettencourt LMA, Lobo J, Helbing D, Kuehnert C,  West GB. Growth,. Innovation, Scaling, and the Pace of Life in Cities,
{\it Proc Natl Acad Sci USA} {\bf 104}, 7301  (2007)






%18
\bibitem{santo}  Chatterjee A, Mitrovi{\v c} M, Fortunato S. {\it Scientific Reports} {\bf 3}, 1049 (2013)
%\bibitem{votes12} C. Borghesi, J-C. Rynal, J-P. Bouchaud, PLOSone, 7(5), e36289 (2012).

%19
\bibitem{firms} Axtell RL. Zipf Distribution of U.S. Firm Sizes, {\it Science} {\bf 293}, 1818 (2001)

%20
\bibitem{1sta} Kemeny J, Snell JL  {\it Mathematical Models in the Social Sciences} (MIT Press, Cambridge, Mass. 1978).

%21
\bibitem{oppi} Castellano C,Fortunato S,Loreto V. Statistical physics of social dynamics, {\it Rev Mod Phys}, {\bf 81}, 591  (2009)

%\bibitem{ciudad2} Marsil M, Yi-Cheng Zhang (1998) Interacting Individuals Leading to Zipf's Law, {\it Phys Rev Lett} 80:2741.

%\bibitem{net2} Newman MEJ, Barabasi AL, Watts DJ (2006) \emph{The Structure and Dynamics of Complex Networks} (Princeton University Press, Princeton).

%22
\bibitem{nosEPJB10} Hernando A et al. Unravelling the size distribution of social groups with information theory in complex networks, \emph{Eur Phys J B} {\bf 76}, 87  (2010)




%23
\bibitem{gmodel} Krings G, et al. Urban gravity: a model for inter-city telecommunication flows {\it J Stat Mech} L07003 (2009)

%24
\bibitem{mob1} Gonz\'alez MC, Hidalgo CA, Barab\'asi AL. Understanding individual human mobility patterns. {\it Nature} {\bf 453}, 779  (2008)

%25
\bibitem{nosEPJB12b} Hernando A, Plastino A, Plastino AR. MaxEnt and dynamical information, \emph{Eur Phys J B} {\bf 85}, 147 (2012)

%26
\bibitem{nosPA10} Hernando A, Vesperinas C, Plastino A. Fisher information and the thermodynamics of scale-invariant systems,{\it  Physica A}, {\bf 389}, 490  (2010)

%27
\bibitem{nosPRE12} Hernando A, Plastino A. The thermodynamics of urban population flows, \emph{Phys Rev E} {\bf 86}, 066105  (2012)


%29
\bibitem{census} Census bureau website, Government of USA, www.census.gov.

%31
\bibitem{biv}  Weisstein, Eric W. Bivariate Normal Distribution. From MathWorld-A Wolfram Web Resource. (mathworld.wolfram.com/BivariateNormalDistribution.html)

%\bibitem{mob2} Simini F et al. (2012) A universal model for mobility and migration patterns, {\it Nature}, 484:96.


%\bibitem{nosPLA12} Hernando A, Plastino A (2013) Scale-invariance underlying the logistic equation and its social applications, \emph{Phys Lett A}, 377:176.

%\bibitem{nosEPJB12} Hernando A, Plastino A (2012) Variational principle underlying scale invariant social systems, \emph{Eur Phys J B} 85:293.


%\bibitem{pnas3} Plane DA, Henrie CJ, Perry MJ (2005) Migration up and down the urban hierarchy and across the life course,{\it Proc Natl Acad Sci USA} 102(43):15313-15318.

%30
\bibitem{tools2} DEMIFER - Demographic and Migratory Flows Affecting European Regions and Cities. ESPON. (www.espon.eu/main/Menu\_Projects/Menu\_Applied Research/demifer.html)

\bibitem{map} Wikimedia Commons (commons.wikimedia.org/wiki/File:USA\_Counties\_with\_FIPS\_and\_names.svg)


\end{thebibliography}
\end{document}